\documentclass[prd,preprint,showpacs]{revtex4}
\usepackage{graphicx}
\usepackage{latexsym}
\usepackage{amsmath}
\begin{document}
\title{Observational constraints on the acceleration of the Universe}
\author{Yungui Gong}
\email{gongyg@cqupt.edu.cn}
\affiliation{College of Electronic Engineering, Chongqing
University of Posts and Telecommunications, Chongqing 400065,
China}
\affiliation{CASPER, Physics Department, Baylor University,
Waco, TX 76798, USA}
\author{Anzhong Wang}
\email{anzhong_wang@baylor.edu}
\affiliation{CASPER,
Physics Department, Baylor University, Waco, TX 76798, USA}
\begin{abstract}

We propose a new parametrization of the deceleration parameter to
study its time-variation behavior. The advantage of parameterizing
the deceleration parameter is that we do not need to assume any
underlying theory of gravity. By fitting the model to the 157 gold
sample supernova Ia data, we find strong evidence that the
Universe is currently accelerating and it accelerated in the past.
By fitting the model to the 115 nearby and Supernova Legacy Survey
supernova Ia data, the evidence that the Universe is currently
accelerating is weak, although there is still a strong evidence
that the Universe once accelerated in the past. The results obtained from
the 157 gold sample supernova Ia data and those from the 115 supernova Ia data
are not directly comparable because the two different data sets measure the
luminosity distance up to different redshifts.

We  then use the Friedmann
equation and a dark energy parametrization to discuss
the same problem. When we fit the model to the supernova Ia data alone, we find
weak evidence that the Universe is accelerating and the current matter density is higher
than that measured from other experiments. After we add the Sloan Digital
Sky Survey data to constrain the dark energy model, we find that the behavior of the
deceleration parameter is almost the same as that obtained from
parameterizing the deceleration parameter.
\end{abstract}
\preprint{astro-ph/0601453}
\maketitle

\section{Introduction}

Astronomical observations suggest the existence of dark energy
which has negative pressure and contributes about 2/3 of the
critical density to the total matter density of the Universe
within the framework of Einstein's general relativity
\cite{sp99,sdss,wmap}. While a wide variety of
dynamical dark energy models were proposed in the literature
\cite{DE}, there are also model-independent studies on the nature
of dark energy by using the observational data. In particular, one usually parameterizes dark
energy density or the equation of state parameter $w(z)$ of dark
energy \cite{par1,par2,par3,par4,par5,gong05}. For the parametrization of $w(z)$,
we need to determine $\Omega_{m0}$ in addition to the parameters in $w(z)$. Furthermore,
one needs to assume the validity of general relativity in all these studies.

Since supernova
(SN) Ia data measures the luminosity distance redshift relationship, it provides a purely
kinematic record of the expansion history of the Universe. It is possible to probe
the evolution of the Hubble parameter or the deceleration parameter by using SN Ia data
without assuming the nature and evolution of the dark energy \cite{TS02,daly,riess}.
For instance,
by parameterizing the deceleration parameter $q(z) = q_{0} + q_{1} z$,
Riess {\em et al} \cite{riess} were able
to study the kinematics of the universe. However, it was soon realized that
such a parametrization cannot re-produce the behavior of the cosmological constant \cite{Virey05}.
An alternative parametrization is a piecewise constant acceleration with two distinct epochs \cite{TS02},
for which it was found that SN Ia data favors recent acceleration and
past deceleration. Recently Shapiro and Turner (ST) applied several simple parametrization
of the deceleration parameter $q(z)$ to the 157 SN Ia data \cite{riess} to
study the property of $q(z)$ \cite{sturner}. They found that
{\em there is little or no evidence
that the Universe is presently accelerating, and that  there is very strong evidence that the
Universe once accelerated}. The advantage of parameterizing $q(z)$ is that the conclusion
does not depend on any particular gravitational theory. The disadvantage is that
it will not give much direct information on the cause of an accelerated Universe.

The conclusion arrived in \cite{sturner} was based on a simple three epoch model of $q(z)$, in which
the function $q(z)$ is not smooth. Since the current SN Ia data is still
sparse, the division of the data to three different redshift bins may not be a good
representation of the data. Therefore, the conclusion based
on this technique needs to be further studied.
Following ST, we propose a simple smooth function of $q(z)$
which we believe is more realistic and then apply the observational
data to get the behavior of the deceleration parameter. After we are sure that the
Universe experienced acceleration, we assume that the acceleration is due
to the presence of dark energy and use a simple dark energy parametrization to study
the property of dark energy in the framework of general relativity.

Specifically, the paper is organized as follows. In section II,
we study the property of $q(z)$ by fitting the parametrization $q(z)=1/2+(q_1 z+q_2)/(1+z)^2$
to the 157 SN Ia data and
the 115 nearby SN Ia and the Supernova Legacy Survey (SNLS) SN Ia data
compiled in \cite{astier}. In section III, we apply the parametrization $w(z)=w_0+w_1 z/(1+z)^2$
to study the property
of $q(z)$ and the nature of dark energy. The Sloan Digital Sky Survey (SDSS) \cite{sdss}
and the Wilkinson Microwave Anisotropy Probe (WMAP) data \cite{wmap} are also combined with the SN Ia data
in our analysis. In section IV, we conclude the paper with some discussion.

\section{The current acceleration}
\begin{figure}
\centering
\includegraphics[width=12cm]{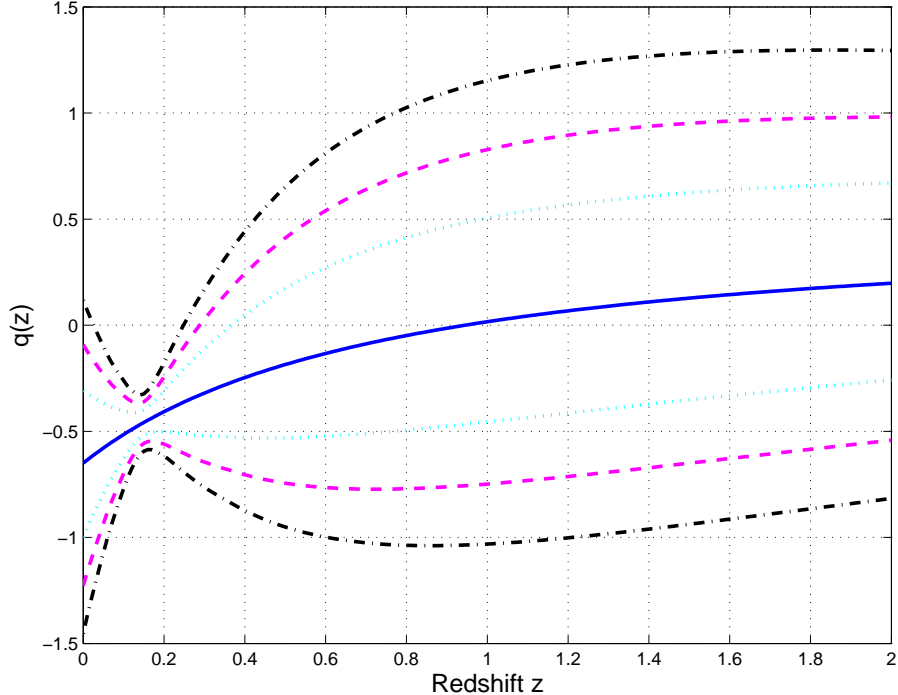}
\caption{The evolution of $q(z)=1/2+(q_1 z+q_2)/(1+z)^2$ by fitting it to the 115 SN Ia data.
The solid line is drawn by using the best fit parameters. The dotted lines
show the $1\sigma$ error, the dashed lines show the $2\sigma$ error and the dotted dash lines show the
$3\sigma$ error.} \label{fig6}
\end{figure}

\begin{figure}
\centering
\includegraphics[width=12cm]{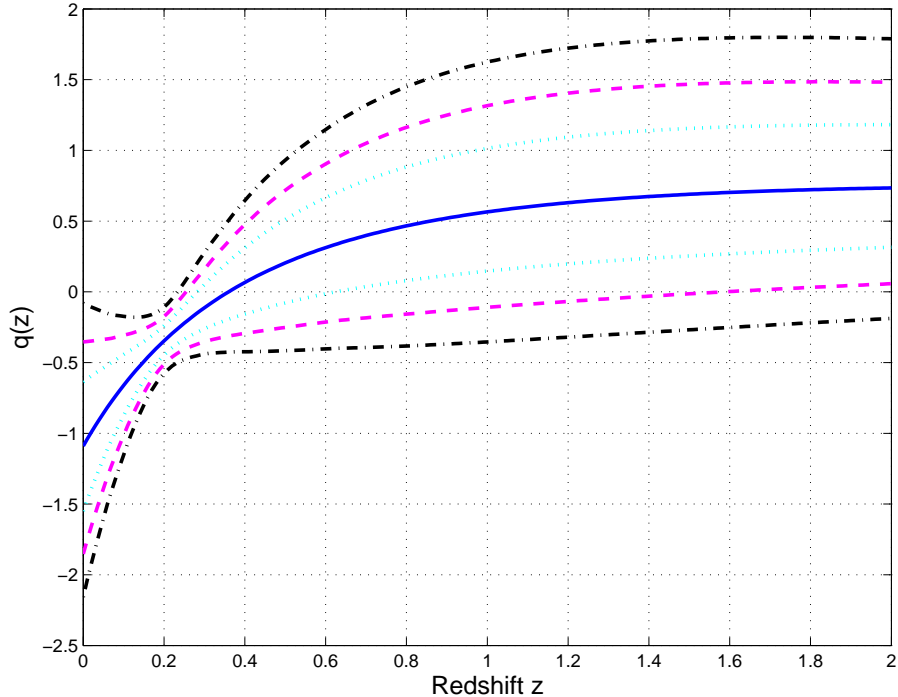}
\caption{The evolution of $q(z)=1/2+(q_1 z+q_2)/(1+z)^2$ by fitting it to the 157 gold sample SN Ia data.
The solid line is drawn by using the best fit parameters. The dotted lines
show the $1\sigma$ error, the dashed lines show the $2\sigma$ error and the dotted dash lines show the
$3\sigma$ error.} \label{fig7}
\end{figure}

From the definition of the Hubble constant $H(t)=\dot{a}/a$ and the deceleration
parameter $q(t)=-\ddot{a}/(aH^2)$, we have
\begin{equation}
\label{hubq}
H(z)=H_0\exp \left[\int^z_0 [1+q(u)]d\ln(1+u)\right],
\end{equation}
where the subscript 0 means the current value of the variable.
So if we are given a function of $q(z)$, then we can find the evolution of our Universe without
applying any particular theory of gravity. ST considered a simple
three-epoch model \cite{sturner}
\begin{equation}
\label{qmodl}
q(z)=\begin{cases}
q_0 & \text{for $z\le z_t$},\\
q_1 & \text{for $z_t<z< z_e=0.3$},\\
q_2=0.5 & \text{for $z\ge z_e=0.3$}.
\end{cases}
\end{equation}
By using Eq. (\ref{hubq}), for the simple three-epoch model
we get
\begin{equation}
\label{hubsl1}
H(z)=\begin{cases}
H_0(1+z)^{1+q_0} & \text{for $z\le z_t$},\\
H_0(1+z_t)^{1+q_0}\left(\frac{1+z}{1+z_t}\right)^{1+q_1} & \text{for $z_t<z< z_e$},\\
H_0(1+z_t)^{1+q_0}\left(\frac{1+z_e}{1+z_t}\right)^{1+q_1}\left(\frac{1+z}{1+z_e}\right)^{1+q_2} &
\text{for $z\ge z_e$}.
\end{cases}
\end{equation}
The parameters in the model are determined by minimizing
\begin{equation}
\label{chi}
\chi^2=\sum_i\frac{[\mu_{obs}(z_i)-\mu(z_i)]^2}{\sigma^2_i},
\end{equation}
where the extinction-corrected distance modulus $\mu(z)=5\log_{10}[d_L(z)/{\rm Mpc}]+25$,
$\sigma_i$ is the total uncertainty in the SN Ia data,
and the luminosity distance is
\begin{equation}
\label{lum}
d_L(z)=(1+z)\int_0^z\frac{dz'}{H(z')}.
\end{equation}
ST found that a long epoch of deceleration is consistent with the 157
gold sample SN Ia data at the 10\% level, and they concluded that there is little
or no evidence that the Universe is presently accelerating.
Although this model is very simple, the functions $q(z)$ and $H(z)$ take
different forms at different epoches. Especially, $q$ is not continuous. Because
the current SN Ia data is still sparse and the decomposition to only three
redshift bins is not a good representation, the conclusion derived from
the model may not be robust. In this paper, we would like to use both the
157 gold sample SN Ia data and the 115 nearby
SN Ia and the SNLS SN Ia data compiled in \cite{astier}. In this 115 data set,
there is no SN Ia with redshift $0.101<z<0.249$ which is around the redshift $z_t$ in
Eq. (\ref{qmodl}).
Therefore it is not a good idea to fit the simple three-epoch model to the 115 SN Ia data, so we proposed
a simple two-parameter function
\begin{equation}
\label{qmod2}
q(z)=\frac{1}{2}+\frac{q_1 z+q_2}{(1+z)^2},
\end{equation}
to fit the 115 SN Ia data. Note that $q(z)\rightarrow 1/2$ when $z\gg 1$ and $q_0=1/2+q_2$,
so the parameter $q_2$ gives the value of $q_0$.
The behavior of $q(z)$ in this parametrization is quite general except that
$q(z)\rightarrow 1/2$ when $z\gg 1$ which is consistent with observations.
If $q_1>0$ and $q_2>0$, then there is no acceleration at all. It is
also possible that the Universe has been accelerating since some time in the past. If $q_1<0$ and $q_2>-1/2$,
then it is possible that the Universe is decelerating and has past acceleration and deceleration.
In other words, this
model may have the same behavior as the simple three-epoch model. The values
of $q_1$ and $q_2$ and the behavior of $q(z)$ can be obtained by fitting
the model to the observational data. Substitute Eq. (\ref{qmod2}) into Eq. (\ref{hubq}),
we get
\begin{equation}
\label{hubsl2}
H(z)=H_0(1+z)^{3/2}\exp\left[\frac{q_2}{2}+\frac{q_1 z^2-q_2}{2(1+z)^2}\right].
\end{equation}
Fitting the model to the 115 SN Ia data, we get $\chi^2=113.65$, $q_1=-0.8^{+2.3}_{-2.2}$
and $q_2=-1.15^{+0.34}_{-0.35}$, here the given error is the $1\sigma$ error.
By using the best fitting results, we plot the evolution of $q(z)$ in Fig. \ref{fig6}.
From Fig. \ref{fig6}, we see that $q_0<0$, i.e., the Universe is currently accelerating,
contains over 96\% of the
probability. At the $3\sigma$ level, it
is possible that $q_0>0$. So it is possible that the Universe is decelerating now although
the evidence is not strong. It seems that we have strong evidence that
the Universe had acceleration in the recent past. We see that the transition redshift when the
Universe underwent the transition from deceleration to acceleration is $z_t=0.95^{+3.25}_{-0.58}$
at the $1\sigma$ confidence level.
However, the redshift range
for past deceleration is around 0.2 which is in the same range
that there is no SN Ia data. To make a more solid conclusion, we also fit
the model to the 157 gold sample SN Ia data and found that $\chi^2=174.07$, $q_1=1.85^{+2.23}_{-2.11}$
and $q_2=-1.59^{+0.45}_{-0.46}$. We plot the evolution of $q(z)$ by using the fitting results in Fig.
\ref{fig7}.

From Fig. \ref{fig7}, we see that
$q(z)<0$ for $0\le z \alt 0.2$ contains over 99.7\% of the probability. This result suggests that there
are strong evidence that the Universe is currently accelerating and the Universe once accelerated.
Fig. \ref{fig7} shows that the transition redshift $z_t=0.36^{+0.24}_{-0.08}$.

\section{Dark Energy Parametrization}
\begin{figure}
\centering
\includegraphics[width=12cm]{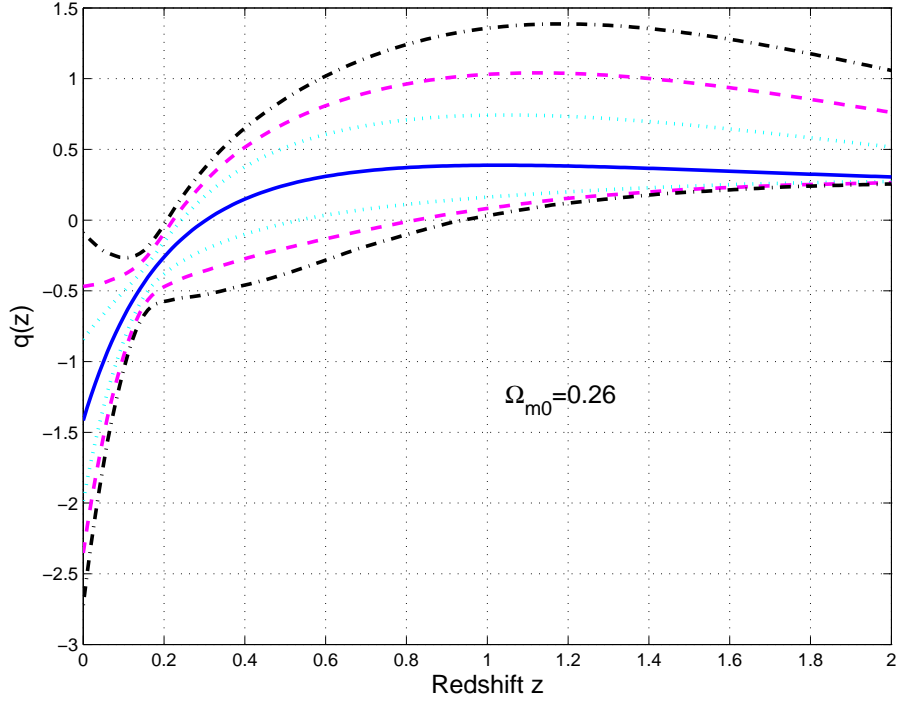}
\caption{The evolution of $q(z)$ by fitting the model $w(z)=w_0+w_1 z/(1+z)^2$ to the
combined 157 gold sample SN Ia and SDSS data.
The solid line is drawn by using the best fit parameters. The dotted lines
show the $1\sigma$ error, the dashed lines show the $2\sigma$ error and the dotted dash lines show the
$3\sigma$ error.} \label{fig9}
\end{figure}

Although the above simple model fits the data well and it tells us that the Universe experienced
acceleration, but it hardly tells us anything about the property of dark energy. In
order to make connection with dark energy, we apply Einstein's general relativity
and work the problem in the usual way. We work on the dark energy parametrization
\begin{equation}
\label{wzeq}
w(z)=w_0+\frac{w_1 z}{(1+z)^2},
\end{equation}
because $w_0$ may be positive and $w(z)\rightarrow w_0$ when $z\gg 1$. This parametrization
may give a currently decelerating universe. It is valuable to mention that the above parametrization
is just a Taylor expansion in the scale factor $a(t)$ to the second order,
$w(z)=w_0+w_1 (a/a_0)-w_1 (a/a_0)^2$, so it has limitations too \cite{bassett}.
The dimensionless dark energy density is
\begin{equation}
\label{deneq}
\Omega_{DE}(z)=\Omega_{DE0}(1+z)^{3(1+w_0)}\exp\left[3w_1 z^2/2(1+z)^2\right].
\end{equation}
Therefore, we get
\begin{equation}
\label{qzeq1}
q(z)=\frac{\Omega_{m0}+(1-\Omega_{m0})(1+z)^{3w_0}[1+3w_0+3w_1 z/(1+z)^2]\exp[3w_1 z^2/2(1+z)^2]}{2[
\Omega_{m0}+(1-\Omega_{m0})(1+z)^{3w_0}\exp[3w_1 z^2/2(1+z)^2]]}.
\end{equation}
So $q_0=(\Omega_{m0}+(1-\Omega_{m0})(1+3w_0))/2$. When $w_0=0$, we get $q_0=0.5>0$. When $w_0=-0.4$,
we can get $q_0>0$ for reasonable value of $\Omega_{m0}$. The model was used to fit the combined 157 gold sample
SN Ia , the SDSS and the WMAP data in \cite{gong05}. Unfortunately, there was an sign mistake in $w_a$
in the results. We fit the model again to the combined 157 gold sample SN Ia and SDSS data and we find that
$\chi^2=172.84$, $\Omega_{m0}=0.26^{+0.05}_{-0.04}$, $w_0=-1.72^{+0.72}_{-0.66}$ and $w_1=6.23^{+4.87}_{-5.72}$.
Combining the contour data of $w_0$ and $w_1$ with Eq. (\ref{qzeq1}),
we plot the evolution of $q(z)$ in
Fig. \ref{fig9}. From Fig. \ref{fig9}, it is evident that $q(z)<0$ for $0\le z \alt 0.2$
contains over 99.7\% of the probability. Fig. \ref{fig9} shows that $z_t=0.30^{+0.23}_{-0.06}$.

Now let us fit the model to the 115 SN Ia data alone, we get
the best fitting values $\chi^2=113.36$, $\Omega_{m0}=0.39^{+0.14}_{-0.39}$, $w_0=-0.97^{+1.31}_{-1.00}$
and $w_1=-4.4^{+9.1}_{-28.5}$. So $w_0=0$ is within the $1\sigma$ error of the best fitting result, which means
that for the fitting to the 115 SN Ia data alone, it is possible that the Universe is experiencing deceleration now.
If we fit the model to the 157 gold sample SN Ia data alone, the result remains the same. In
fact, if we set $w_0=0$ and fit the 115 SN Ia data, we get the best
fitting values $\chi^2=115.53$, $\Omega_{m0}=0.46^{+0.07}_{-0.09}$ and $w_1=-21.9^{+7.9}_{-15.9}$. The result is
consistent with the result of setting $w_0$ free. Note that the best fitting value of $\Omega_{m0}$ is higher than that
obtained from other observations. To fix this problem, we add the SDSS data to the SN Ia data. So we add the term
$(A-0.469)^2/0.017^2$ to Eq. (\ref{chi}). The parameter $A=0.469\pm 0.017$ measured from the SDSS data is defined as
\begin{equation}
\label{paraa}
A=\frac{H_0\sqrt{\Omega_{m0}}}{0.35}\left[\frac{0.35}{H(0.35)}\left(\int^{0.35}_0\frac{dz}{H(z)}\right)^2\right]^{1/3}.
\end{equation}

\begin{figure}
\centering
\includegraphics[width=12cm]{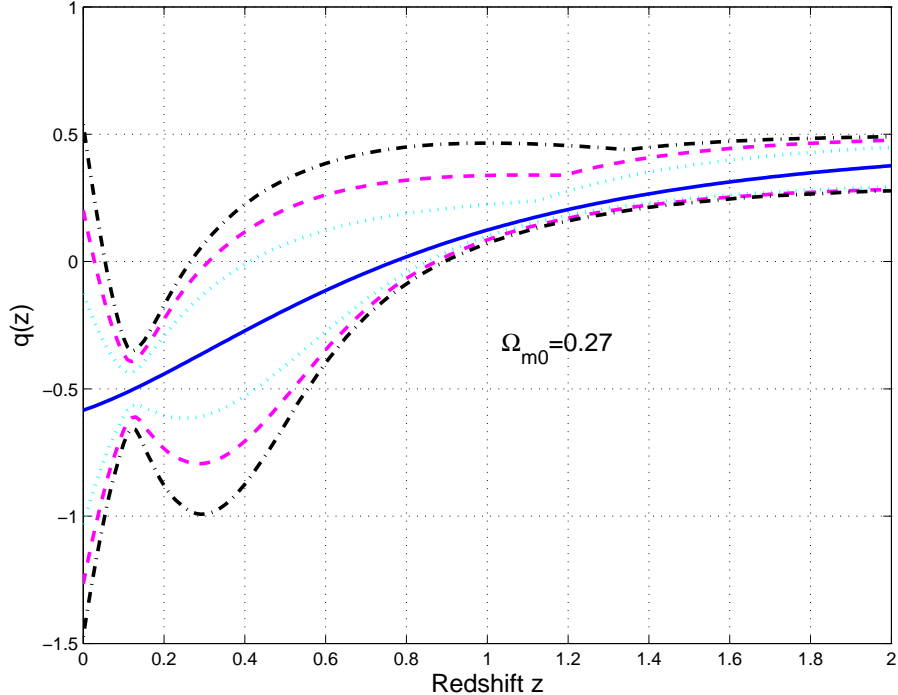}
\caption{The evolution of $q(z)$ by fitting the model $w(z)=w_0+w_1 z/(1+z)^2$ to the
combined 115 SN Ia and SDSS data.
The solid line is drawn by using the best fit parameters. The dotted lines
show the $1\sigma$ error, the dashed lines show the $2\sigma$ error and the dotted dash lines show the
$3\sigma$ error.} \label{fig10}
\end{figure}

The best fitting results to the combined 115 SN Ia and SDSS data for the model (\ref{wzeq}) are
$\chi^2=113.55$, $\Omega_{m0}=0.27\pm 0.04$, $w_0=-0.99^{+0.58}_{-0.54}$ and $w_1=-0.30^{+5.11}_{-5.95}$.
Combining the contour data of $w_0$ and $w_1$ with Eq. (\ref{qzeq1}), we plot the evolution of $q(z)$ in
Fig. \ref{fig10}. From Fig. \ref{fig10}, we see that the evidence that $q_0<0$ is weak and there is
strong evidence that the Universe once accelerated. Fig. \ref{fig10} shows that $z_t=0.77^{+0.08}_{-0.36}$.
Of course, we may add the shift
parameter derived from WMAP data to fit the model. By adding WMAP data, we found that the conclusion remains the same.
We also find that $w_0=-0.4$ is within $2\sigma$ error and $w_0=0$ is within $3\sigma$ error.
For $w_0=-0.4$, we find the best fitting results are $\chi^2=117.13$, $\Omega_{m0}=0.28\pm 0.03$
and $w_1=-6.2^{+1.0}_{-1.1}$. For $w_0=0$, we find the best fitting results are $\chi^2=122.80$,
$\Omega_{m0}=0.29\pm 0.03$
and $w_1=-10.4^{+1.1}_{-1.3}$. These results are also summarized in table 1.

\begin{table}
\caption{Summary of the best fit parameters to the combined 115 SN Ia and SDSS data.}
\label{fittab}
\begin{center}
\begin{tabular}{|c|c|c|c|c|c|c|}
  \hline
$\Omega_{m0}$&$w_0$&$w_1$&$q_0$&$\chi^2$\\
  \hline
$0.27\pm 0.04$&$-0.99^{+0.58}_{-0.54}$&$-0.30^{+5.11}_{-5.95}$&-0.58&113.55\\\hline
$0.28\pm 0.03$&$-0.4$&$-6.2^{+1.0}_{-1.1}$&0.068&117.13\\\hline
$0.29\pm 0.03$&0&$-10.4^{+1.1}_{-1.3}$&0.5&122.80\\\hline
\end{tabular}
\end{center}
\end{table}

\begin{figure}
\centering
\includegraphics[width=12cm]{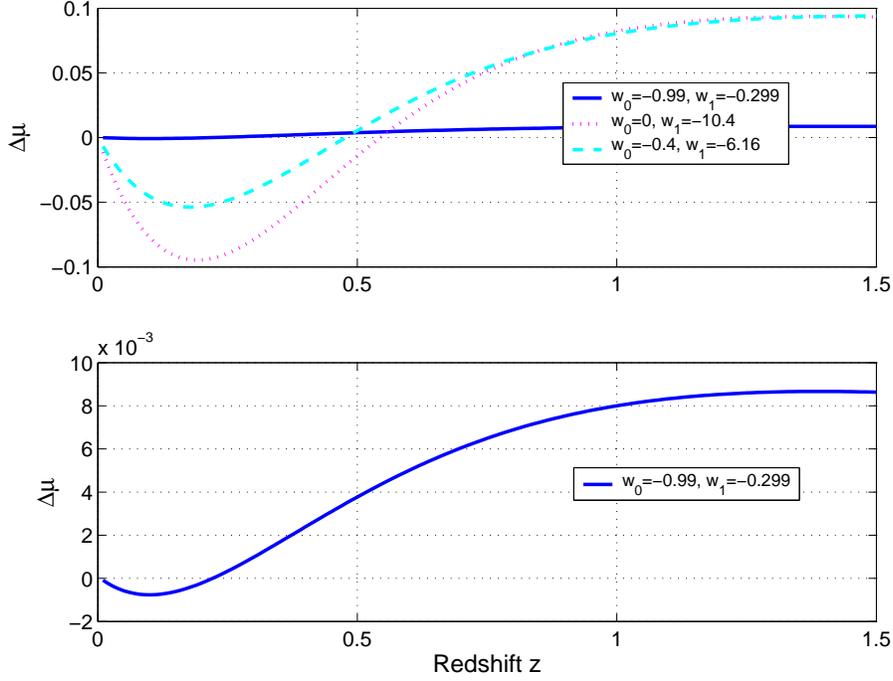}
\caption{The dependence of relative magnitude-redshift relation upon parameters $w_0$ and $w_1$,
relative to the $\Lambda$CDM model with $\Omega_{m0}=0.27$. All parameters are the best fit
values to the combined 115 SN and SDSS data. The bottom panel plots the curve $w_0=-0.99$ and $w_1=-0.299$
again so that the scale can be seen more clearly.} \label{fig4}
\end{figure}

In Fig. \ref{fig4}, we plot the relative magnitude for the three cases discussed above with respect to
the $\Lambda$CDM model with $\Omega_{m0}=0.27$. The higher the redshift, the bigger the difference. While
it is easier to distinguish the current deceleration models with the $\Lambda$CDM model with more accurate observational
data, it is more difficult to distinguish the general dynamical dark energy model with the $\Lambda$CDM model.

\section{Discussion}
By fitting the parametrization $q(z)=1/2+(q_1 z+q_2)/(1+z)^2$ to the 157 gold sample SN Ia data,
we find that $q(z)<0$ for $0\le z \alt 0.2$ within the uncertainty of $3\sigma$ level. Recall that
$q(z)<0$ means that the Universe is in the acceleration phase at redshift $z$. We also
get $z_t=0.95^{+3.25}_{-0.58}$ at the $1\sigma$ level. If we
fit the parametrization to the 115 nearby and SNLS SN Ia data, we find that $q_0>0$ at the $3\sigma$
level and strong evidence that the Universe once accelerated. We also find
that $z_t=0.36^{+0.24}_{-0.08}$ at the $1\sigma$ level. Note that these conclusions do not
depend on any particular theory of gravity.

To study the role that general relativity might play, in this paper we also
considered the Friedmann equation together with  the dark energy parametrization
$w(z)=w_0+w_1 z/(1+z)^2$. After applying them to
the 157 gold sample SN Ia data or the 115 SN Ia data, we find that $q_0>0$ at the $1\sigma$ level and
the best fitting $\Omega_{m0}\sim 0.4$, which is higher than that determined by other observations. In
the parametrization $q(z)$, we have two parameters $q_1$ and $q_2$. When we
use the dark energy parametrization, there are three parameters $\Omega_{m0}$,
$w_0$ and $w_1$. With the addition of one more parameter apart from the
assumption that the acceleration is due to dark energy in the framework of general relativity,
we expect that the constraints
will be loose. Therefore the
dark energy parametrization is then fitted to the combined SN Ia and SDSS data because
the combined data will break the degeneracies among the three parameters. If we fit the dark energy
model to the combined 157 gold sample SN Ia and SDSS data, we find that $q(z)<0$ for $0\le z \alt 0.2$
contains over 99.7\% of the probability. We also get $z_t=0.30^{+0.23}_{-0.06}$ at the $1\sigma$
level. If we fit the dark energy
model to the combined 115 SN Ia and SDSS data, we find that $q_0>0$ at the $2\sigma$
level and strong evidence that the Universe once accelerated.
The transition redshift is found to be $z_t=0.77^{+0.08}_{-0.36}$ at the $1\sigma$ level. Whether
we use the parametrization $q(z)$ or the dark energy parametrization $w(z)$ to fit the
SN Ia data, the evidence that the the Universe once accelerated is very strong. These results confirm the
conclusion that there is strong evidence that the Universe once accelerated obtained in \cite{sturner}.
Our results also suggest that $z_t\agt 0.2$. Although the upper bound on $z_t$ is too large
by using the parametrization $q(z)$, $z_t$ is more tightly constrained to be $z_t\alt 1.0$ by
using the dark energy parametrization $w(z)$.

The situation about $q_0$ is more subtle. It depends on the model and the data. If we fit the models to the 157
gold sample SN Ia data, we find strong evidence for $q_0<0$. But if the models are fitted to the 115
SN Ia data, the conclusion is different. The discrepancy is caused by the difference
of the redshift range the data probed.
The 157 gold sample SN Ia data probes much deeper in the redshift ($z\sim 1.7$) than the 115 SNLS data
which stops at $z\sim 1$. For the dark energy model, the evidence for current deceleration
looks promising if the model is fitted to the SN Ia data alone. The fitting also
gives larger value for $\Omega_{m0}$.
When the SN Ia data is combined with the SDSS data for the dark energy model, we get reasonable value for
$\Omega_{m0}$ and the sign of $q_0$ is uncertain. The uncertainty of the question
whether the Universe in accelerating now is directly related to the value of $w_0$.
If $w_0\lesssim -0.6$, then the Universe is accelerating currently. If $w_0<-1$, then
the dark energy behaves like a phantom and it is not explained by the quintessence field.
In fact, different models mean different external
priors, so it shows that whether the Universe is currently accelerating depends on external prior.

In conclusion, we confirm that there is strong evidence that the Universe once accelerated.

\begin{acknowledgments}
Y.G. Gong is supported by Baylor University, NNSFC under grant No. 10447008 and 10575140,
SRF for ROCS, State Education Ministry
and CQUPT under grant No. A2004-05.
\end{acknowledgments}



\begin{thebibliography}{nbound}
\bibitem{sp99} S. Perlmutter {\it et al.},
Astrophy. J. {\bf 517}, 565 (1999); P.M. Garnavich  {\it et al.}, Astrophys. J. {\bf 493}, L53 (1998);
A.G. Riess  {\it et al.}, Astron. J. {\bf 116}, 1009 (1998).
\bibitem{sdss} D.J. Eisenstein {\it et al.}, Astrophys. J. {\bf 633}, 560
(2005).
\bibitem{wmap} C.L. Bennett {\it et al.}, Astrophys. J. Supp. Ser. {\bf 148}, 1
(2003).
\bibitem{DE} V. Sahni  and A. A. Starobinsky, Int. J. Mod. Phys. D
{\bf 9}, 373 (2000); T. Padmanabhan, Phys. Rep.
{\bf 380}, 235 (2003); P.J.E. Peebles and B. Ratra, Rev. Mod. Phys. {\bf 75}, 559 (2003);
V. Sahni, {\it The Physics of the Early Universe}, P. 141, edited by E. Papantonopoulos (Springer, 2005);
T. Padmanabhan, astro-ph/0510492.
\bibitem{par1} M. Chevallier and  D. Polarski, Int. J. Mod.
Phys. D {\bf 10}, 213 (2001);
E.V. Linder, Phys. Rev. Lett. {\bf 90}, 091301
(2003);
T.R. Choudhury and  T. Padmanabhan, Astron. Astrophys. {\bf 429}, 807 (2005);
B. Feng, X.L. Wang and X.M.
Zhang, Phys. Lett. B {\bf 607}, 35 (2005);
H.K. Jassal, J.S. Bagla and T. Padmanabhan,
Mon. Not. Roy. Astron. Soc. {\bf 356}, L11 (2005).
\bibitem{par2} J. Weller  and A. Albrecht, Phys. Rev. Lett. {\bf
86}, 1939 (2001);
D. Huterer and M.S. Turner, Phys. Rev. D {\bf 64}, 123527 (2001);
J. Weller  and A. Albrecht, Phys. Rev. D {\bf 65}, 103512 (2002).
P. Astier, Phys. Lett. B {\bf 500}, 8 (2001).
\bibitem{par3} G. Efstathiou, Mon. Not. Roy. Soc. {\bf 310},
842 (1999);
B.F. Gerke and G. Efstathiou, Mon. Not. Roy. Soc. {\bf
335}, 33 (2002); P.S. Corasaniti and E.J. Copeland, Phys. Rev. D {\bf
67}, 063521 (2003); C. Wetterich, Phys. Lett. B {\bf 594}, 17 (2004);
S. Lee, Phys. Rev. D {\bf 71}, 123528 (2005);
K. Ichikawa and T. Takahashi, astro-ph/0511821.
\bibitem{par4} U. Alam, V. Sahni, T.D. Saini and A.A.
Starobinsky, Mon. Not. Roy. Astron. Soc. {\bf 354}, 275 (2004);
U. Alam, V. Sahni and A.A.
Starobinsky, J. Cosmol. Astropart. Phys. JCAP {\bf 0406} (2004) 008;
Y.G. Gong, Int. J. Mod. Phys. D {\bf 14} (2005) 599; Y.G. Gong, Class. Quantum Grav. {\bf 22}, 2121 (2005).
\bibitem{par5} J. J\"{o}nsson, A. Goobar, R. Amanullah and L.
Bergstr\"{o}m, J. Cosmol. Astropart. Phys., JCAP {\bf 0409} (2004) 007; Y. Wang and P. Mukherjee,
Astrophys. J. {\bf 606}, 654 (2004); Y. Wang and M. Tegmark, Phys.
Rev. Lett. {\bf 92}, 241302 (2004); V.F. Cardone, A. Troisi and S. Capozziello,
Phys. Rev. D {\bf 69}, 083517 (2004); D. Huterer and A. Cooray, Phys. Rev. D {\bf 71}, 023506 (2005).
\bibitem{gong05} Y.G. Gong and Y.Z. Zhang, Phys. Rev. D {\bf 72}, 043518 (2005).
\bibitem{TS02} M.S. Turner and A.G. Riess {\it et al.}, Astrophys. J. {\bf 569}, 18 (2002).
\bibitem{daly} R.A. Daly and S.G. Djorgovski, Astrophys. J. {\bf 597}, 9
(2003); R.A. Daly and S.G. Djorgovski, Astrophys. J. {\bf 612}, 652 (2004).
\bibitem{riess} A.G. Riess {\it et al.}, Astrophys. J. {\bf 607}, 665 (2004).
\bibitem{Virey05}   J.-M. Virey {\it et al.}, Phys. Rev.
{\bf D72},  061302 (2005).
\bibitem{sturner} C.A. Shapiro and M.S. Turner, astro-ph/0512586.
\bibitem{astier} P. Astier {\it et al.}, astro-ph/0510447.
\bibitem{bassett} B.A. Bassett, P.S. Corasaniti and M. Kunz, Astrophys. J. {\bf 617}, L1 (2004).

\end{thebibliography}
\end{document}